\documentclass[prl,twocolumn,superscriptaddress,amsmath,amssymb,showpacs,floatfix,preprintnumbers]{revtex4}
\usepackage{amsmath}
\usepackage{graphicx}
\usepackage{dcolumn}
\usepackage{color}
\DeclareGraphicsExtensions{.png .jpg .pdf}

\begin{document}

\title{Bidirectional switching assisted by interlayer exchange coupling in asymmetric magnetic tunnel junctions}

\author{D. J. P. de Sousa}\email{sousa020@umn.edu}
\affiliation{Department of Electrical and Computer Engineering, University of Minnesota, Minneapolis, Minnesota 55455, USA}
\author{P. M. Haney}
\affiliation{Physical Measurement Laboratory, National Institute of Standards and Technology,
Gaithersburg, Maryland 20899-6202, USA}
\author{D. L. Zhang}
\author{J. P. Wang}\email{jpwang@umn.edu}
\author{Tony Low}\email{tlow@umn.edu}
\affiliation{Department of Electrical and Computer Engineering, University of Minnesota, Minneapolis, Minnesota 55455, USA}

\date{ \today }

\begin{abstract}
We study the combined effects of spin transfer torque, voltage modulation of interlayer exchange coupling and magnetic anisotropy on the switching behavior of perpendicular magnetic tunnel junctions (p-MTJs). In asymmetric p-MTJs, a linear-in-voltage dependence of interlayer exchange coupling enables the effective perpendicular anisotropy barrier to be lowered for both voltage polarities. This mechanism is shown to reduce the critical switching current and effective activation energy. Finally, we analyze the possibility of having switching via interlayer exchange coupling only.

\end{abstract}


\maketitle

Magnetic tunnel junctions (MTJs) that can be switched bidirectionally by electrical means are highly desirable for low power consumption applications\cite{ref1}. Current-induced magnetization reversal is one of the most promising and reliable technologies available for achieving bidirectional switching in MTJs\cite{ref2, ref3, ref4, ref5}. 

\begin{figure}[t]
\centerline{\includegraphics[width = \linewidth]{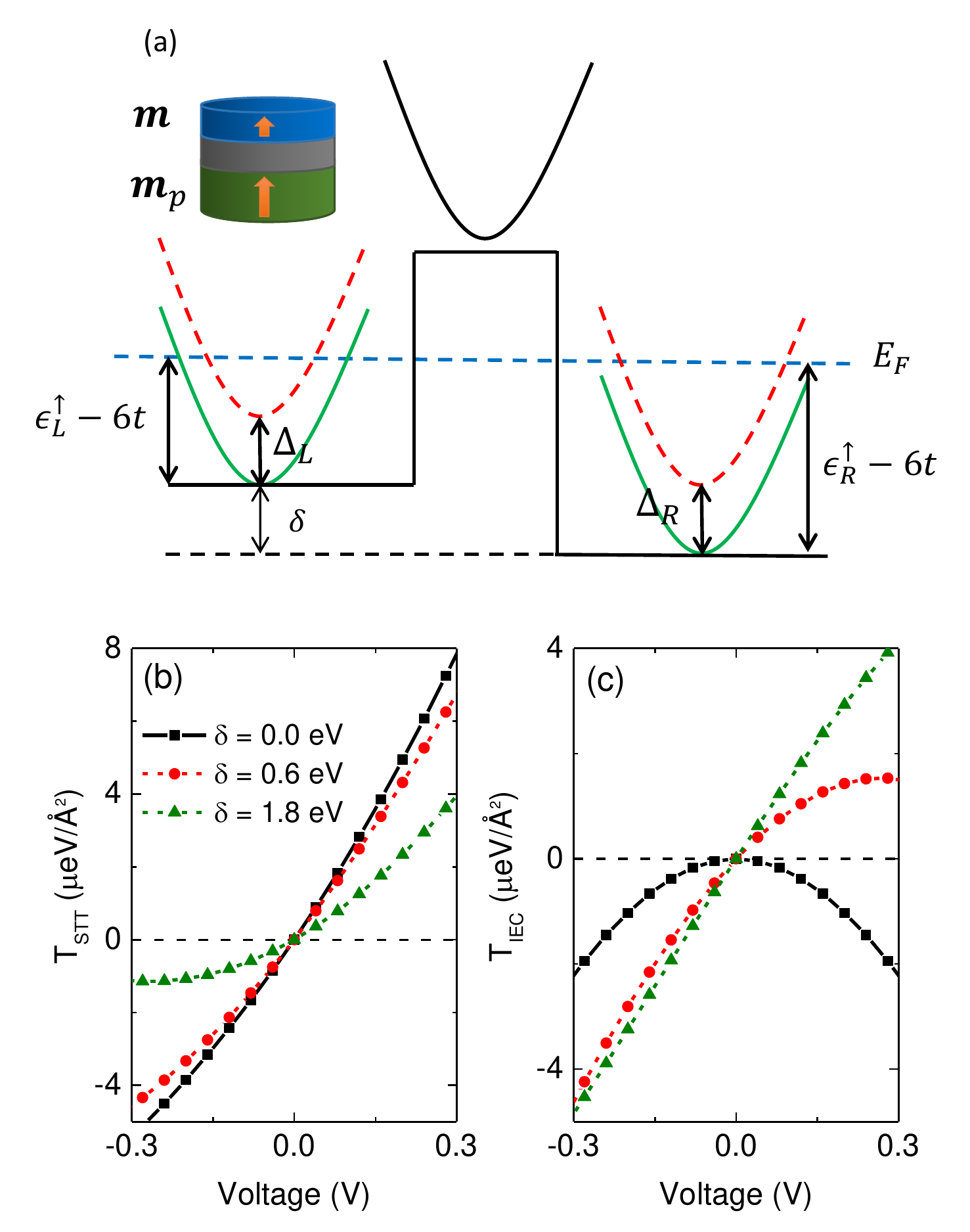}}
\caption{(Color online) (a) MTJ band diagram. The parameter $\delta = \epsilon_{L}^{\uparrow(\downarrow)} - \epsilon_{R}^{\uparrow(\downarrow)}$ controls the asymmetry of the MTJ. The bottom of the spin up (down) bands in the single orbital tight-binding approach is $\epsilon^{\uparrow(\downarrow)} - 6t$, where $t$ is the nearest neighbor hopping parameter. The inset shows a sketch of an asymmetric p-MTJ with $\textbf{m}$ and $\textbf{m}_p$ corresponding to the unit vectors in the direction of the magnetization of the free and fixed layer, respectively. Panels (b) and (c) show the voltage dependence of spin transfer torque and non-equilibrium interlayer exchange coupling, respectively, for different MTJ asymmetries $\delta$, as defined in (a).}
\label{Fig1}
\end{figure}

Current-induced switching relies on spin transfer torque (STT), where the interaction between current-carrying spins which are misaligned with the magnetization leads to magnetic dynamics and reversal\cite{ref4}.  In addition to STT, a charge current modifies the interlayer exchange coupling (IEC) between fixed and free layers via an additional field-like torque\cite{ref5, Haney, ref11}. Though frequently called "field-like spin transfer torque component", in this work we refer to this torque component as interlayer exchange torque\cite{Haney}. Denoting the free (pinned) magnetic layer orientation by $\textbf{m}$ ($\textbf{m}_p$) [See inset Fig.~\ref{Fig1}(a)], the total current-induced torque density is
\begin{eqnarray}
\mathcal{N} = T_{\textrm{IEC}} \textbf{m}\times \textbf{m}_p + T_{\textrm{STT}}\textbf{m}\times( \textbf{m}\times \textbf{m}_p).
\label{eq0}
\end{eqnarray}

Unlike in spin valves where the IEC torque is negligible, it has been demonstrated that $T_{\textrm{IEC}}$ is comparable to $T_{\textrm{STT}}$ in MgO-based MTJs, considerably affecting the magnetization dynamics of the free layer\cite{Kubota, Sankey, Deac}. However, while the importance of $T_{\textrm{STT}}$ for magnetization switching is well understood, the contribution of $T_{\textrm{IEC}}$ is often omitted in many analyses and poorly explored.

For perpendicular MTJs (p-MTJs), the critical switching current $J_c$ is directly proportional to the total effective perpendicular anisotropy $K_{\textrm{eff}}$\cite{ref6, ref7}. Such proportionality reflects the fundamental problem encountered in memory technology, where one seeks to improve $K_{\textrm{eff}}$ for better retention of information while reducing the critical switching current $J_c$ for low-power consumption\cite{ref7}. Particularly, the voltage control of magnetic anisotropy (VCMA) is currently being quoted as one of the most promising methods to circumvent this problem, as it provides a mechanism to reduce the anisotropy barrier $K_{\textrm{eff}} \mathcal{V}$, where $ \mathcal{V}$ is the volume of the free layer, only when a voltage is applied across the MTJ, enabling one to reduce $J_c$ momentarily while maintaining a sizeable $K_{\textrm{eff}}$ at zero applied voltage\cite{ref7, ref8, ref9, ref10}. However, while it can reduce the critical switching current for a given applied voltage by reducing $K_{\textrm{eff}}$, it tends to increase $K_{\textrm{eff}}$ for the opposite voltage polarity. The ability to overcome the anisotropy barrier bidirectionally while decreasing the critical current density is highly desirable, and remains a long-standing goal in the search for low power consumption spintronics. 

In this work, we show that $T_{\textrm{IEC}}$ can assist STT switching by effectively reducing the anisotropy barrier for both voltage polarities in asymmetric p-MTJs. We demonstrate that $T_{\textrm{IEC}}$ directly competes with the total effective intrinsic uniaxial anisotropy quantified by $K_{\textrm{eff}}$, enabling one to reduce the critical switching current bidirectionally by tuning the degree of asymmetry of the p-MTJ. Our model includes the combined effects of STT, VCMA and IEC effects on p-MTJs, which are all known to be present in this kind of system\cite{ref7, refLLG}.

The total torque acting on the magnetization vector of the free layer is decomposed into different contributions, as given by the Landau-Lifshitz-Gilbert (LLG) equation\cite{ref13_2}
\begin{eqnarray}
\frac{d \textbf{m}}{dt} =  -\gamma \textbf{m}\times\textbf{H}_{\textrm{eff}} + \alpha \textbf{m}\times \frac{d\textbf{m}}{dt}  + \frac{\gamma}{\mu_0 M_S t_{\textrm{free}}}\mathcal{N},
\label{eq1}
\end{eqnarray}
where $\textbf{m} = \textbf{M}/M_S$, with $\textbf{M}$ being the magnetization of the free layer with saturation $M_S$, $\gamma $ is the gyromagnetic ratio, $\alpha$ is the intrinsic damping parameter, $\mu_0$ is the vacuum permeability and $t_{\textrm{free}}$ is the thickness of the free layer. The effective field is $\textbf{H}_{\textrm{eff}} = (2K_{\textrm{eff}}(V)m_z/\mu_0 M_S )\textbf{z}$, with $\textbf{z}$ being the axis perpendicular to the free layer plane and $m_z$ being the $z$ component of $\textbf{m}$. The total effective anisotropy coefficient is given by $K_{\textrm{eff}}(V) = K_{\textrm{eff}}(0) + \xi V$ with $K_{\textrm{eff}}(0)= K_{i}/t_{\textrm{free}} - \mu_0 M_S^2 /2$ being the effective perpendicular magnetic anisotropy at zero voltage with interfacial anisotropy $K_i$. The VCMA coefficient is $\xi$ and $V$ the applied voltage across the p-MTJ. We assume $\textbf{m}_p = \textbf{z}$, i.e., perpendicular to the interface.

The critical switching voltage $V_c$ is given by the following implicit equation [See Supplementary information]
\begin{eqnarray}
& T_{\textrm{STT}}(V_c) = 2\alpha t_{\textrm{free}} \left(K_{\textrm{eff}}(V_c)m_z - \frac{T_{\textrm{IEC}}(V_c)}{2t_{\textrm{free}}}\right),
\label{eq2}
\end{eqnarray}
where $m_z = \pm 1$ for magnetization initially in the parallel (P, with $m_z = +1$) or antiparallel (AP, with $m_z = -1$) configuration. This result reveals that while $T_{\textrm{STT}}$ acts in favor or against the intrinsic damping\cite{ref4}, $T_{\textrm{IEC}}$ competes directly with the anisotropy torque, affecting the final critical STT switching magnitude $T_{\textrm{STT}}^c = T_{\textrm{STT}}(V_c)$. Before analyzing the consequences of this equation from the perspective of the quantum transport model, let's suppose, for simplicity, the following voltage dependencies of the torques, i.e., $T_{\textrm{STT}} = \beta_{\textrm{STT}}V$, $T_{\textrm{IEC}} = C_1 V + C_2 V^2$, where the coefficients $\beta_{\textrm{STT}}$, $C_1$ and $C_2$ express the voltage modulation of the non-equilibrium torques to lowest order in V. Our convention for the voltage is that $V > 0$ leads to an electron flow from the fixed layer to the free layer. For symmetric p-MTJs, $T_{\textrm{IEC}}$ is an even function of applied voltage, i.e., the spatial top-bottom symmetry requires that $C_1 = 0$ and $C_2 \neq 0$\cite{Ioannis}. In this case, one can solve Eq.~(\ref{eq2}) analytically for $V_c$ to find
\begin{eqnarray}
& V_c = \frac{2\alpha t_{\textrm{free}}}{\beta_{\textrm{STT}}} K_{\textrm{eff}}(0) m_z,
\label{eq3}
\end{eqnarray}
where we have assumed $\xi = 0$, i.e., no VCMA effect, and neglected terms of order $\alpha^2$. Interestingly, Eq.~(\ref{eq3}) shows that $V_c$ does not depend on $C_2$ in this limit. Hence, this result is consistent with the fact that $T_{\textrm{IEC}}$ has little or no influence on the magnetization switching in conventional symmetric p-MTJs.

The situation for asymmetric p-MTJs is different. In this case, theoretical\cite{ref11, ref12} and experimental\cite{ref13} analysis have shown that $C_1 \neq 0$, giving a sizable linear voltage-dependent contribution to $T_{\textrm{IEC}}$. In this situation, $T_{\textrm{IEC}}$ acts like a torque due to an effective field with sign determined by $V$ and direction aligned with the magnetization of the fixed layer. For a given applied voltage $V$, this results in an unidirectional anisotropy, to be contrasted with the intrinsic uniaxial magnetic anisotropy. 
We explore the consequences of this symmetry breaking induced contribution by assuming, for simplicity, $\xi = 0$ and $C_2 = 0$. Equation~(\ref{eq2}) can then be easily solved:
\begin{eqnarray}
& V_c = \frac{2\alpha t_{\textrm{free}}}{\beta_{\textrm{STT}}} K_{\textrm{eff}}(0)m_z\left(1 + \alpha C_1 / \beta_{\textrm{STT}}\right)^{-1},
\label{eq4}
\end{eqnarray}
where $V_c$ is reduced by a factor of $1 + \alpha C_1/\beta_{\textrm{STT}}$. This simple analysis shows the relevance of $T_{\textrm{IEC}}$ in reducing the critical switching current. A comparison between experiments from Refs.~\cite{Kubota, ref13} indicates that $C_1 = 0$ and $C_1 \approx 30 $ kA/m for symmetric and asymmetric CoFeB/MgO/CoFeB MTJs, respectively. These results show the possibility of tuning $V_c$ via $C_1$.

The above analysis, albeit qualitative, demonstrates the possibility of reducing critical switching voltage when $T_{\textrm{IEC}}$ exhibits strong asymmetric dependence on voltage, i.e., $C_1 \gg 0$. According to Eq.~(\ref{eq4}), the sign of $C_1 / \beta_{\textrm{STT}}$ must be positive in order to decrease the $V_c$. While $\beta_{\textrm{STT}}$ is usually positive, it was experimentally observed that one can tune the sign and magnitude of $C_1$ by controlling the relative composition between fixed and free magnetic layers\cite{ref13}. In the following section, we first evaluate the voltage modulation of both $T_{\textrm{STT}}$ and $T_{\textrm{IEC}}$ within a single orbital quantum transport model and explore the dependence of the critical current density with p-MTJ asymmetry. 

\emph{Non-equilibrium torques}. In the absence of spin-orbit coupling\cite{Paul}, the torque exerted  on the magnetization of the \textit{i}-th atomic plane of the free layer is related to the spin current flux into that plane as $\textbf{T}_i = -\nabla \cdot \textbf{Q}_i =  \textbf{Q}_{i-1,i} - \textbf{Q}_{i,i+1}$ where $\textbf{Q}_{i,j}$ is the spin-current density between atomic planes $i$ and $j$. The total torque exerted on the semi-infinite magnetic lead reads $\textbf{T} = \sum_{i} \textbf{T}_i = \textbf{Q}_{\textrm{Ox}/\textrm{FM}}$, where $\textbf{Q}_{\textrm{Ox}/\textrm{FM}}$ is the spin-current density penetrating the magnetic lead at the oxide-ferromagnet interface\cite{ref11, ref12}. Assuming a spin quantization axis along the $\textbf{m}_p = \textbf{z}$ direction for the fixed layer, the $T_{\textrm{STT}}$ and $T_{\textrm{IEC}}$ components are obtained by extracting the $\textbf{m}\times( \textbf{m}\times \textbf{m}_p)$ and $\textbf{m}\times \textbf{m}_p$ components, respectively, of the interface spin-current $\textbf{Q}_{\textrm{Ox}/\textrm{FM}}$\cite{Ioannis, ref11, ref12}.

We employ the single-orbital tight-binding model and express the spin-current density as\cite{Ioannis, ref11, ref12}
\begin{eqnarray}
\textbf{Q}_{i,j} = \frac{1}{4\pi} \int_{\Omega_B} \frac{d^2 \textbf{k}_{||}}{(2\pi)^2}\int dE\operatorname{Tr}_{\sigma}[(H_{ji}G_{ij}^{<}-H_{ij}G_{ji}^{<})\vec{\sigma}],
\label{eq5}
\end{eqnarray}
where $\vec{\sigma} = (\sigma_x, \sigma_y, \sigma_z)$ is the vector of Pauli matrices, $H_{ij}$ is hopping matrix between sites $i$ and $j$, $G_{ij}^{<}$ is the lesser Green's function of the whole coupled system and the $\textbf{k}_{||}$ integration is performed over the 2D in-plane Brillouin zone $\Omega_B$. This model provides an accurate description of the voltage dependence of the non-equilibrium torques in systems such as in Fe/MgO/Fe MTJs\cite{ref14, ref15, ref16, ref17}. 

\begin{figure}[t]
\centerline{\includegraphics[width = \linewidth]{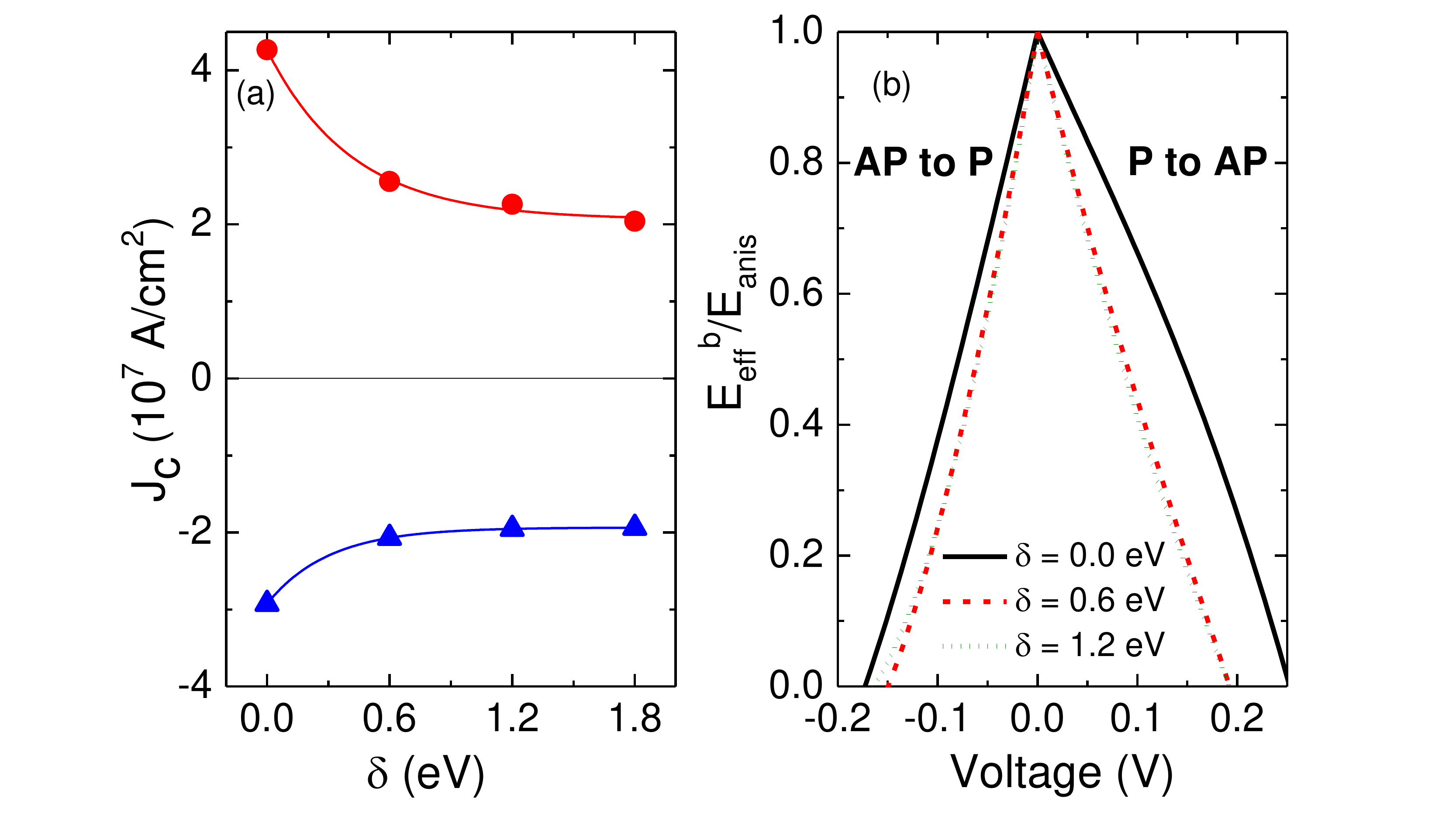}}
\caption{(Color online) Critical current density as function of the asymmetry parameter $\delta$. The red circles (blue triangles) show the trend for P to AP (AP to P) switching. (b) Normalized effective activation energy as function of applied voltage for different p-MTJ asymmetries  for P to AP ($V > 0$) and AP to P ($V < 0$) switching.}
\label{Fig3}
\end{figure} 

In experiments, asymmetry in the ferromagnetic contacts can be introduced through the use of different metals\cite{Miller}, or by considering ferromagnets with different compositions such as in Co$_{40}$Fe$_{40}$B$_{20}$/MgO/Co$_{49}$Fe$_{21}$B$_{20}$ MTJs\cite{ref13}. In this work, we introduce asymmetry in the ferromagnets by adjusting their band fillings. The symmetry breaking is controlled by the asymmetry parameter $\delta = \epsilon_{R}^{\uparrow(\downarrow)} - \epsilon_{L}^{\uparrow(\downarrow)}$, where $\epsilon_{L(R)}^{\uparrow(\downarrow)}$ refers to the spin-up (down) band filling of the left (right) magnetic lead, as shown by the band diagram in Fig.~\ref{Fig1}(a). The exchange splitting inside the ferromagnets are kept constant and the same, i.e., $\Delta_L = \Delta_R$. 

The voltage dependence of $T_{\textrm{STT}}$ and $T_{\textrm{IEC}}$ for different asymmetries ($\delta = 0.0$ eV (solid black), $\delta = 0.6$ eV (dashed red) and $\delta = 1.8$ eV (dot-dashed olive)) are shown in Fig.~\ref{Fig1}(b) and (c), respectively. The angular dependencies of both torque components are $\sin(\theta)$. Hence, it suffices to show only their amplitudes. The results of Fig.~\ref{Fig1}(b) show that $T_{\textrm{STT}}$ presents an approximately linear behavior for small applied voltages, i.e., $T_{\textrm{STT}} \approx \beta_{\textrm{STT}}V$, with a slope that decreases as an increasing function of the asymmetry parameter $\delta$. In particular, for the most asymmetric case considered ($\delta = 1.8$ eV), the voltage behavior of $T_{\textrm{STT}}$ at negative $V$ deviates from linear and one can potentially achieve $T_{\textrm{STT}}$ sign reversal under applied voltages for one of the polarities\cite{Ioannis}. Figure.~\ref{Fig1}(c) indicates that $T_{\textrm{IEC}}$ is quadratic in $V$ for symmetric p-MTJs, i.e $T_{\textrm{IEC}} \approx C_2 V^2$ with $C_2 < 0$, as theoretically predicted and observed experimentally \cite{ref4, ref11, ref12, Kubota}. As one increases the asymmetry via $\delta$, the voltage modulation of $T_{\textrm{IEC}}$ is enhanced while an additional linear-in-voltage contribution develops, i.e $T_{\textrm{IEC}} \approx C_1 V + \mathcal{O}(V^2)$. We also emphasize that the ratio $C_1/\beta_{\textrm{STT}}$ is positive if one choose $\delta > 0$. 

\emph{Critical current density.} The critical current density $J_c$ is computed by computing the current-voltage relation using Landauer's formula and non-equilibrium Green's function.  This relation enables the previously computed voltage-dependent $T_{\textrm{IEC}}$ and $T_{\textrm{SST}}$ to be converted to their corresponding current-dependent. Figure~\ref{Fig3}(a) shows $J_c$ as a function of the asymmetry parameter $\delta$ for P to AP (red circles) and AP to P (blue triangles) switching with $K_i \approx 1.3$ mJ/m$^2$, and VCMA coefficient $\xi = 20 $ kJ/V$\cdot$m$^3$. The result clearly shows that $J_c$ decreases with asymmetry, which can be interpreted as follows: The presence of $T_{\textrm{IEC}}$ in asymmetric p-MTJs reduces $K_{\textrm{eff}}$ for both voltage polarities, as qualitatively described by Eq.~(\ref{eq4}). Therefore, the effective energy barrier between P and AP configurations decreases and less current is necessary for magnetization reversal.

The symmetry breaking also has important consequences for thermally activated switching. Following Ref.~\cite{ref19}, we have derived expressions for the effective activation energy in the presence of $T_{\textrm{IEC}}$:
\begin{eqnarray}
&E_{eff}^b = E_{anis}\left(1 - \frac{T_{\textrm{STT}}}{T_{\textrm{STT}}^c}\right)\left(1 - m_z\frac{T_{\textrm{IEC}}}{2t_{\textrm{free}}K_{\textrm{eff}}}\right),
\label{eq5}
\end{eqnarray}
from which one can extract the switching time $\tau^{-1} = f_0 \exp(-E_{\textrm{eff}}^b /k_B T)$, with $f_0$ being an attempt frequency. The anisotropy energy barrier $E_{\textrm{anis}}(V) = K_{\textrm{eff}}(V)\mathcal{V}$ quantifies the thermal stability factor $\Delta = E_{\textrm{anis}}(0)/k_B T$. Figure~\ref{Fig3}(b) shows the voltage dependence of the normalized effective energy barrier $E_{\textrm{eff}}^b / E_{\textrm{anis}}$ for P to AP ($V > 0$) and AP to P ($V < 0$) switching considering several different asymmetry parameters $\delta$. In this plot we use the voltage dependence of non-equilibrium torques from the quantum transport model. As one can see, the activation energy drops faster with $V$ for asymmetric p-MTJs, allowing for higher switching probabilities at a given temperature $T$.

\textit{Switching by voltage control of IEC}.
So far, we have shown that in asymmetric p-MTJ, $T_{\textrm{IEC}}$ can assist STT switching by effectively reducing the anisotropy barrier for both voltage polarities. Anisotropy and voltage dependent IEC torques can be written as derivatives of an effective energy, given by:
\begin{eqnarray}
&E(\theta) = K_{\rm{eff}}\sin(\theta)^2 + (T_{\rm{IEC}}/t_{\rm{free}})\cos(\theta),
\label{eq7}
\end{eqnarray}
where $\theta$ is the angle between $\textbf{m}$ and $\textbf{m}_p$. Stable equilibrium points are found at energy minima, where the total field-like torque vanishes.

\begin{figure}[t]
\centerline{\includegraphics[width = \linewidth]{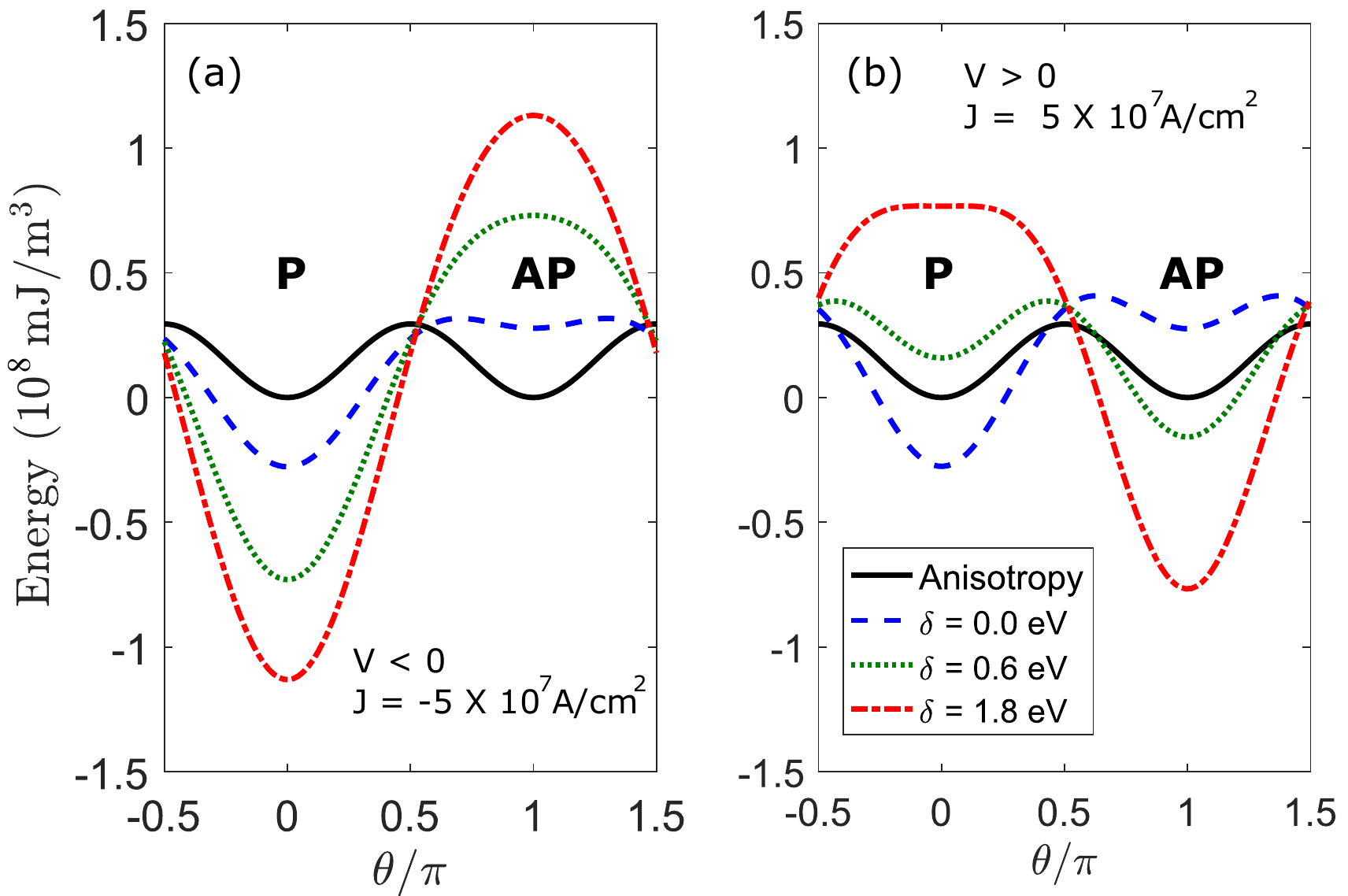}}
\caption{(Color online) Energy landscapes for (a) negative and (b) positive current densities of absolute value $5 \times 10^7$ A/cm$^2$ for different degrees of asymmetries $\delta$. We define $\theta$ as the angle between the magnetization of free and pinned layers such that parallel and anti-parallel configuration, highlighted as P and AP, are found in $\theta = 0$ and $\theta = \pi$, respectively. We considered $K_{\textrm{eff}}(0) = 29.5$ kJ/m$^3$ and $\xi = 20$ kJ/(V $\cdot$ m$^3$). The solid black curve shows contribution of perpendicular anisotropy only, whereas the other curves show the total energy landscape resulting from the sum of IEC and VCMA contributions.}
\label{Fig2}
\end{figure}

Figures \ref{Fig2}(a) and (b) show the energy landscape for negative and positive current-density of $J = 5 \times 10^7$ A/cm$^2$ for different p-MTJ asymmetries. We have also plotted the energy at zero applied voltage in black solid lines for comparison purposes. One sees that $K_{\rm{eff}}$ alone gives rise to two metastable equilibrium configurations with P ($\theta = 0$) or AP ($\theta = \pi$) alignment, emphasizing the axial nature of perpendicular anisotropy. 

Figure~\ref{Fig2}(a) shows the angular dependence of the total energy for different asymmetries $\delta$ at applied $V < 0$. For the symmetric case ($\delta = 0$ eV), a negative bias voltage gives rise to a negative $T_{\textrm{IEC}}$ [See Fig.~\ref{Fig1}(c)] while decreasing $K_{\textrm{eff}}$. The associated energy landscape for this case is shown as a dashed blue curve in Fig.~\ref{Fig2}(a).  One sees that the stability of the P (AP) configuration is enhanced (suppressed) due to the unidirectional nature of the IEC torque. The dotted olive curve in Fig.~\ref{Fig2}(a) shows the angular dependence of energy for the same current density considering an asymmetric p-MTJ with $\delta = 0.6$ eV. In this case, the previously metastable AP configuration is now a maximum, indicating a current-induced instability and subsequent switching from $\theta = \pi$ to $\theta=0$. The dash-dotted red curve shows that the effect is even more pronounced if one further increases the asymmetry to $\delta = 1.8$ eV. 

For $V>0$, $K_{\textrm{eff}}$ now increases with $V$. For symmetric p-MTJs, $T_{\textrm{IEC}}$ is an even function of the bias and, therefore, remains negative with positive applied voltage [See Fig.~\ref{Fig1}(c)]. The resulting energy landscape is represented by the dashed blue curve in Fig.~\ref{Fig2}(b), where one observes an even greater stability in the P configuration, increasing the difficulty to switch from P to AP. In asymmetric p-MTJs, however, $T_{\textrm{IEC}}$ changes sign under reversal of the voltage polarity. Such behavior results in the curves corresponding to $\delta \neq 0$ eV in Fig.~\ref{Fig2}(b). In these cases, the P (AP) configuration tends to become more unstable (stable) as one increases the asymmetry, favoring P to AP switching. In particular, the case $\delta = 1.8$ eV shows that one can completely destabilize the P configuration, showing pure IEC bidirectional bipolar switching.

\emph{Conclusion.} 
We have studied the simultaneous impact of VCMA, IEC and STT for p-MTJs. We demonstrated that for asymmetric devices, linearly varying $T_{\textrm{IEC}}$ plays an important role in STT switching by renormalizing the effective anisotropy barrier. Such effect leads to reduced critical switching current for magnetization reversal, and can even lead to switching based on IEC alone. 

\textit{Acknowledgments}. This work was partially supported by C-SPIN/STARnet, DARPA ERI FRANC program and ASCENT/JUMP.

\end{document}